\documentstyle [12pt,aasms4]{article}
\def\beq{\begin{equation}}
\def\eeq{\end{equation}}
\def\bey{\begin{eqnarray}}
\def\eey{\end{eqnarray}}

\def\mpc{\,h^{-1}{\rm {Mpc}}}
\def\mpci{\,h {\rm {Mpc}}^{-1}}
\def\kms{\,{\rm {km\, s^{-1}}}}

\def\xiz#1{\xi_z(r_{p#1},\pi_{#1})}

\def\gs{\mathrel{\raise1.16pt\hbox{$>$}\kern-7.0pt
\lower3.06pt\hbox{{$\scriptstyle \sim$}}}}
\def\ls{\mathrel{\raise1.16pt\hbox{$<$}\kern-7.0pt
\lower3.06pt\hbox{{$\scriptstyle \sim$}}}}
\def\gtsima{$\; \buildrel > \over \sim \;$}
\def\ltsima{$\; \buildrel < \over \sim \;$}
\def\prosima{$\; \buildrel \propto \over \sim \;$}
\def\gsim{\lower.5ex\hbox{\gtsima}}
\def\lsim{\lower.5ex\hbox{\ltsima}}
\def\simgt{\lower.5ex\hbox{\gtsima}}
\def\simlt{\lower.5ex\hbox{\ltsima}}
\def\simpr{\lower.5ex\hbox{\prosima}}

\begin{document}
\title {
The Scaling of the Redshift Power Spectrum: Observations from the Las Campanas
Redshift Survey}
\author {Y.P. Jing$^{1,2,3,4}$, G. B\"orner$^{1,4,5}$} 
\affil{$ ^1$
Shanghai Astronomical Observatory, the Partner Group of MPI f\"ur
Astrophysik, Nandan Road 80, Shanghai 200030, China}
\affil {$^2$ National Astronomical Observatories, Chinese Academy of
Sciences, Beijing 100012, China}
\affil{$ ^3$ National Astronomical Observatory, Mitaka,
 Tokyo 181-8588, Japan}
\affil{$ ^4$
Research Center for the Early Universe, School of Science, University
of Tokyo, Bunkyo-ku, Tokyo 113, Japan}
\affil {$ ^5$
Max-Planck-Institut f\"ur Astrophysik, Karl-Schwarzschild-Strasse 1,
85748 Garching, Germany}
\affil {e-mail: ypjing@center.shao.ac.cn, ~grb@mpa-garching.mpg.de}
\received{---------------}
\accepted{---------------}

\begin{abstract}
In a recent paper we have studied the redshift power spectrum
$P^S(k,\mu)$ in three CDM models with the help of high resolution
simulations. Here we apply the method to the largest available
redshift survey, the Las Campanas Redshift Survey (LCRS). The basic
model is to express $P^S(k,\mu)$ as a product of three factors
\beq
P^S(k,\mu)=P^R(k)(1+\beta\mu^2)^2 D(k,\mu)\,.
\eeq
Here $\mu$ is the cosine of the angle between the wave vector and the
line of sight.  The damping function $D$ for the range of scales
accessible to an accurate analysis of the LCRS is well approximated by
the Lorentz factor
\beq
D=[1+{1\over 2}(k\mu\sigma_{12})^2]^{-1}\,.
\eeq
We have investigated different values for $\beta$ ($\beta=0.4$, 0.5,
0.6), and measured the real space power spectrum $P^R(k)$ and the
pairwise velocity dispersion $\sigma_{12}(k)$ from $P^S(k,\mu)$ for
different values of $\mu$. The velocity dispersion $\sigma_{12}(k)$ is
nearly a constant from $k=0.5$ to 3 $\mpci$. The average value for
this range is $510\pm 70 \kms$.  The power spectrum $P^R(k)$ decreases
with $k$ approximately with $k^{-1.7}$ for $k$ between 0.1 and 4
$\mpci$.
The statistical
significance of the results, and the error bars, are found with the
help of mock samples constructed from a large set of high resolution
simulations.  A flat, low-density ($\Omega_0=0.2$) CDM model can give
a good fit to the data, if a scale-dependent special bias scheme is
used which we have called the cluster-under-weighted bias (Jing et
al.). 

\end{abstract}


\section {Introduction}

The spatial distribution of galaxies can be retrieved approximately,
when the redshift is used as a measure of the distance. It is not
exact, because the galaxies in general deviate from the linear Hubble
flow, but the redshift distortions can provide us with valuable
information about the dynamics of galaxies.  A traditional method for
studying the redshift distortion is the redshift two-point correlation
function (RTPCF) $\xiz{}$ (\cite{gp1973}, \cite{p1980},
\cite{beanetal1983}, \cite{dp1983} \cite{mjb1993},
\cite{marzkeetal1995}, \cite{fisheretal1994},
\cite{ratcliffeetal1998}, \cite{postmanetal1998}, \cite{gg1999},
\cite{smalletal1999}). Assuming certain functional forms for the
distribution function (DF) of the pairwise velocity (say, an
exponential form, \cite{p1976}) and for the average infall velocity
(say, the form of the self-similar solution, \cite{dp1983}), one can
construct a model for $\xiz{}$ which describes the real situation,
when the coupling between the peculiar velocity and the spatial
density of the galaxies is weak (\cite{p1980}). A comparison of the
model with observations for $\xiz{}$ provides a test for the validity
of the model assumptions (such as the DF of the pairwise velocity) and
a determination of the pairwise velocity dispersion
$\sigma_{12}(r)$. Early studies of small redshift samples
(\cite{p1976},\cite{p1979}) have shown that an exponential form for
the DF of the pairwise velocity is preferable to a Gaussian form. This
important conclusion has been confirmed by a number of later studies
based on much larger surveys ( \cite{dp1983}, \cite{marzkeetal1995},
\cite{fisheretal1994}).  The pairwise velocity dispersion was not easy
to pin down; the value sensitively depends on the presence or absence
of rare rich clusters in the redshift surveys which contained only a
few thousand galaxies (\cite{mjb1993}, \cite{zureketal1994},
\cite{marzkeetal1995}). Motivated by this fact, several authors have
since attempted (\cite{kss1997}, \cite{dmw1997}, \cite{soc1998},
\cite{bdl2000}) to find new statistics for the thermal motion of
galaxies which are less sensitive to the regions of rich clusters. The
new statistics, by their design, are more robust with respect to the
sampling of rich clusters than the pair-weighted $\sigma_{12}(r)$ and
have produced interesting constraints on models of galaxy formation,
but the results sometimes are more difficult to interpret in the
context of dynamical theories (e.g. the Cosmic Virial Theorem and the
Cosmic Energy Equation). Fortunately, the pair-weighting will no
longer be a problem for future large redshift surveys like 2dF and
SDSS which will contain hundreds to thousands of rich clusters in the
survey. In fact, the currently largest publicly available redshift
survey, the Las Campanas Redshift Survey (LCRS;
\cite{shectmanetal1996}) which contains about 30 rich clusters, is
already large enough to reduce the sampling error in $\sigma_{12}(r)$
to $15\%$ only (\cite{jmb1998}, hereafter JMB98).  Therefore the
pairwise velocity dispersion, which is a well-defined physical
quantity and has simple relations to the dynamical theories (the
Virial Theorem), will remain an important quantity to measure in
observations.

Although the pair-weighting is not a problem for the statistics of
$\sigma_{12}(r)$ with the largest redshift surveys or the next
generation of redshift surveys, the currently widely used method of
measuring $\sigma_{12}(r)$ (\cite{dp1983}) does have its limitations
(JMB98). Functional forms must be first assumed for the DF of the
pairwise velocity and the infall velocity, and it must be assumed that
the spatial density and peculiar velocity of galaxies are uncorrelated
so that the redshift two-point correlation function (model) can be
written as a convolution of the real-space correlation with the DF of
the pairwise velocity. Although the validity of the functional form
for the DF of the pairwise velocity can be checked by comparing the
model of the RTPCF to the observed one, the tests of different
functional forms are rather limited. The functional form may 
depend on the separation between the objects (\cite{jfs1998},
\cite{mjs2000}). The infall velocity also has a significant effect on the
determination of $\sigma_{12}(r)$, because the two quantities are
rather degenerate in modeling $\xiz{}$ (\cite{jb1998}). Moreover, the
density and velocity are strongly coupled at the small scales where previous
studies have measured $\sigma_{12}(r)$. These limitations can lead to
some systematic bias in the estimation of $\sigma_{12}(r)$ (JMB98; 
\cite{jb1998}). These systematic effects were not so
important in early statistical studies of $\sigma_{12}(r)$ since the
sampling errors in $\sigma_{12}(r)$ were dominant. Since the sampling
error is just 15\% for the LCRS (JMB98) and will be even
smaller for the upcoming surveys, it is important to consider
alternative methods to measure the
redshift distortion and $\sigma_{12}(r)$ which are less model dependent.

In a recent paper (\cite{jb2001}, hereafter JB2001), we have studied the
redshift power spectrum $P^S(k,\mu)$ in three typical CDM cosmological
models, where $\mu$ is the cosine of the angle between the wave
vector and the line-of-sight. Two distinctive biased tracers, the
primordial density peaks (\cite{bardeenetal1986}; \cite{davisetal1985}) and
the cluster-under-weighted population of particles (JMB98),
are considered in addition to the pure dark matter models.  Based on a large
set of high-resolution simulations, we have measured the redshift power
spectrum for these three tracers from the linear to the non-linear
regimes. For all the tracers and the three CDM models, the redshift
power spectrum $P^S(k,\mu)$ can be expressed by
\beq
P^S(k,\mu)=P^R(k) (1+\beta\mu^2)^2 D(k\mu \sigma_{12}(k))
\label{eq1}
\eeq
where $P^R(k)$ is the real space power spectrum, and $\beta =
\Omega^{0.6}/b$ ($\Omega$ is the density parameter and $b$ is the
linear bias parameter). The factor $(1+\beta\mu^2)^2$ accounts for the
linear compression of structures on large scales
(\cite{kaiser1987}). Although $\sigma_{12}(k)$ and the 3-D peculiar
velocity dispersion are not immediately comparable, it was pointed out
in JB2001 that their values are different only by 15\% in the
simulation, if $r=1/k$ is used in the comparison. So $\sigma_{12}(k)$
is a good indicator for the velocity dispersion. In Equation
(\ref{eq1}) the damping function $D$, which should generally depend on
$k$, $\mu$ and $\sigma_{12}(k)$, is a function of one variable $k\mu
\sigma_{12}(k)$ only. The functional form of $D(k\mu \sigma_{12}(k))$
was found to depend on the cosmological model and the bias recipes,
but for small $k$ (large scales) where $D>0.1$, $D(k\mu
\sigma_{12}(k))$ can be accurately expressed by the Lorentz form,
\beq
D(k\mu \sigma_{12}(k))={1\over 1+{1\over 2}k^2\mu^2\sigma^2_{12}(k)}\,.
\label{eq2}
\eeq
This implies that the exponential form is a good model for the
distribution function of the pairwise velocity at large separation.
Applying the model of Equation (\ref{eq1}) to an observation of
$P^S(k,\mu)$ therefore may yield a determination for the four
quantities: $P^R(k)$, $\beta$, $\sigma_{12}(k)$ and the functional
form of $D(y)$ all of which are useful observables for testing galaxy
formation models.  An additional advantage is that the infall
effect is accounted for by the single parameter $\beta$, compared with
a complicated form for the infall velocity used in the correlation
function method.

In this paper, we apply this method to the LCRS, the largest redshift
survey publicly available. The redshift power spectrum $P^S(k,\mu)$ is
measured from the RTPCF $\xiz{}$ of this survey (JMB98) using the
method outlined in Section 2. The result is presented in section
3. The statistical significance of the measurement, and the error
bars, is tested using mock samples in section 4. The final section
(\S5) is devoted to discussion and conclusions.

\section{Method}

We convert the redshift two-point correlation function $\xi_z({\bf s})$ 
to the redshift power spectrum by the Fourier transformation:
\beq
P^S({\bf k})=\int \xi_z({\bf s}) e^{-i{\bf k}\cdot {\bf s}} d{\bf s}\,.
\label{eq3}
\eeq
In cylindrical polar coordinates ($r_p, \phi, \pi$) with the
$\pi$-axis parallel to the line-of-sight, $P^S({\bf k})$ depends on
$k_p$, the wavenumber perpendicular to the line-of-sight, and on
$k_\pi$, the wavenumber parallel to the line-of-sight. The power spectrum
can be written
\beq
P^S(k_p,k_{\pi})=\int \xiz{} e^{-i[k_p r_p\cos(\phi)+\pi k_\pi]}
r_p dr_p d\phi d\pi\,.
\label{eq4}
\eeq
With some elementary mathematical manipulation, we get the following
expression:
\beq
P^S(k_p,k_{\pi})=2\pi \int_{-\infty}^{\infty} d\pi \int_0^{\infty} r_p dr_p
\xiz{} \cos(k_\pi \pi) J_0(k_p r_p)
\label{eq5}
\eeq
where $J_0(k_p r_p)$ is the zeroth-order Bessel function.

We will use the redshift two-point correlation function $\xiz{}$
of the LCRS measured by JMB98. In their
paper, JMB98 analyzed the correlation function for the whole sample as
well as for the three southern slices and the three northern slices
separately. Moreover, they have generated a large sample of mock
surveys with different observational selection effects and measured
$\xiz{}$ for them. Their tests with the mock samples are very
important, as emphasized by JMB98 themselves, for checking the
statistical methods, correcting for the observational selection
effects, and estimating the statistical errors of the measurements. We
will also use their results of the mock samples in this paper for the
same purposes.

In their work, JMB98 measured $\xiz{}$ in equal logarithmic bins of
$r_p$ and in equal linear bins of $\pi$. The reason why different
types of bins are chosen for $r_p$ and $\pi$ is the fact that $\xiz{}$
decreases rapidly with $r_p$ but is flat with $\pi$ on small
scales. Thus this way of presenting $\xiz{}$ is better than using the
$\log$-$\log$ or the linear-linear bins for $r_p$ and $\pi$, and is
also suitable for the present work.  The peculiar velocity of a few
hundred $\kms$ should smoothen out structures on a few $\mpc$ in the
radial direction, and the linear bin of $\Delta \pi_{i}=1\mpc$ is
suitable for resolving the structures in the radial direction. With
logarithmic bins chosen for $r_p$, the $r_p$ dependence is resolved
well, because otherwise the small scale clustering on the projected
direction cannot be recovered.
With this bin method, we obtain the power spectrum:
\beq
P^S(k_p,k_{\pi})=2\pi \sum_{i,j} \Delta \pi_{i} r_{p,j}^2 \Delta \ln r_{p,j}
\xi_z(r_{p,j},\pi_i) \cos(k_\pi \pi_i) J_0(k_p r_{p,j})
\label{eq6}
\eeq where $\pi_{i}$ runs from $-50$ to $50\mpc$ with $\Delta
\pi_{i}=1\mpc$ and $r_{p,j}$ from $0.1$ to $31.6\mpc$ with $\Delta \ln
r_{p,j}=0.288$ (Be careful not to confuse two $\pi$s in
Eqs. (\ref{eq5}) and (\ref{eq6}): the first $\pi$ in the
right-hand-side has the conventional meaning, i.e. 3.14159..., and the
others are for the axis along the line-of-sight.). We use the
summation of Eq.(8) with rectangular boundaries in $\pi$ and
$r_p$. This gives good, unbiased results for both the velocity
dispersion and the power spectrum, as will be seen in \S 4. This works
also better than a spherical window of radially decreasing weighting
which tends to bias the power spectrum estimates. We have tested this
throughly comparing the results for mock samples with those for the
full simulation.

There are several sources which could introduce errors to our
measurement of the redshift power spectrum. One is the Poisson error
related to the discrete sampling of galaxies, and the other is the
cosmic variance related to the limited volume of the survey. Both
types of uncertainty are intrinsic for any redshift survey, and are
already factored into the measured $\xiz{}$. When converting the
correlation function to the power spectrum with Eq.(\ref{eq6}), the finite
bins and the cutoffs of $r_p$ and $\pi$ may cause additional errors.
To reduce the effect of the finite bins, we divide every $\pi$ and
$\ln r_{p}$ bin into $N$ sub-bins, and use the cubic spline method
to interpolate $\xiz{}$. We have run a number of
trials for different numbers of the sub-bins $N$, and found that the
final result is insensitive to $N$. For the results presented in the
following sections, $N=10$ is used. Since these different types of
errors are mixed in a complicated way in the measured power spectrum,
we will use the mock samples to quantify the errors.

\cite{landyetal1998} have also considered the Fourier transform of the
redshift two-point correlation function for the LCRS survey.  Although
the spirit of their work and ours is similar, there are some important
differences. We have included the infall effect in our statistics and
consider the full angular $\mu$-dependence of the redshift power
spectrum. Furthermore, we will use the two dimensional integral
(Eq.\ref{eq5}) to determine both the velocity dispersion and the power
spectrum.

\section {The redshift power spectrum in the Las Campanas redshift survey}

Since the redshift distortion does not change the total number of the
galaxy pairs at a certain projected distance, it is not difficult to
prove from Eq.(3) that $P^S(k,0)=P^R(k)$. Thus the damping factor
$D(k,\mu)$ can be estimated by the following expression
\beq
D(k,\mu)\equiv
{P^S(k,\mu)\over P^S(k,0)(1+\beta \mu^2)^2}\,.
\label{eq7}
\eeq
The quantities on the right hand side can be measured through Eq.(\ref{eq6})
except for the parameter $\beta$. If the observational catalog is
large enough, one can simultaneously determine the damping factor and
$\beta$ as described by JB2001. However, the LCRS is too small
to measure this quantity because, as we can see in Figure 1, there is a large statistical
fluctuation on large scales ($k\sigma_v\mu\ls 1$). This is consistent
with the recent estimates of Matsubara et
al. (\cite{matsubaraetal2000}) based on an eigenmode method, who give
$\beta=0.30\pm 0.39$. Therefore, we will not attempt to determine the
value of $\beta$. Instead we will focus on measuring the damping
function.  Here we fix a value for $\beta$, $\beta=0.5$, which is
consistent with the observations of the cluster abundance
(e.g. \cite{ks1996}) as well as with the statistical result of Matsubara et
al (\cite{matsubaraetal2000}) for the LCRS. We have checked the
results for $\beta=0.4$ and $\beta=0.6$, and found that our
conclusions are little changed if $\beta$ varies within this range.

Figure~1 shows the inverse of the damping function $D(k,\mu)$ we measured
for the Las Campanas redshift survey. Following JB2001, the damping
factor is expressed as a function of the scaling variable $k\mu
\sigma_v$. For this figure, $\sigma_v$ is fixed to be a constant
$ 500\kms$, and only the data points which have a relative
error smaller than $50\%$ are presented. The error of $D(k,\mu)$ is
estimated from the scatter of this quantity between the
cluster-weighted mock samples (see \S 4 about the mock samples).  From
the figure, it is easy to see that our measured result of $D(k,\mu)$
can be well described by a Lorentz form (the solid line). This result
is consistent with the model studies of JB2001 who have shown that the
damping factors of different bias models are all very close to the
Lorentz form when the damping factor $D(k,\mu)\gs 0.1$. The LCRS
is still not large enough to explore the highly non-linear regime
where $D(k,\mu)\ll 0.1$. This also means that an exponential
distribution function for the pairwise velocities is a good
approximation in this range. In addition, we note that there exists
significant scatter on large scales $k\mu\sigma_v<1$, thus it is very
difficult to measure the $\beta$ value accurately with this catalog
(cf. Matsubara et al. \cite{matsubaraetal2000}).

As Figure 1 shows, the damping function that the LCRS can explore is
in the interval $0.1\ls D(k,\mu)< 1$. The accuracy is limited by the
sample-to-sample error (the cosmic variance) on large scale
($D(k,\mu)\approx 1$). Since the LCRS measured the redshift for each
$1.5\times 1.5 {\rm ~deg^2}$ field by one exposure, the dense regions
in the galaxy distribution (e.g. clusters) are under sampled, which
seriously limits the accuracy of our measurement for small
scales [$D(k,\mu)\ll 1$]. For the regime which the LCRS can
effectively explore, the Lorentz form is a good approximation for the
damping function, consistent with the model study of JB2001.
Thus, we can determine the real-space power spectrum $P^R(k)$
and the pairwise velocity dispersion $\sigma_v$ by a least-square
fitting to the observed redshift power spectrum, assuming the Lorentz
form for the damping factor. Although $P^R(k)\equiv P^S(k,0)$
mathematically, for a finite-sized survey like the LCRS, $P^S(k,0)$
fluctuates around the true value of $P^R(k)$ for the limited number of
the modes. We can better determine $P^R(k)$ if we combine $P^S(k,\mu)$
at different angles $\mu$, thus we treat $P^R(k)$ as a free parameter.

We show such best fitting curves (the solid lines) together with the
measured redshift power spectrum in Figure 2. For the curves from top
to bottom, the wavenumber $k$ is incremented by $\Delta \lg k=0.2$
from $0.25\mpci$ to $2.5\mpci$. As the figure shows, the model we use
for $P^S(k,\mu)$ (i.e. the Lorentz form for the damping function and
equation \ref{eq1}) succeeds in describing the result of the LCRS
data.  Our best fitting values for $\sigma_{12}(k)$ and $P^R(k)$ are
plotted in Figures 3 and 4 respectively.  Here we have corrected the
{\it fiber collision effect of the survey} (\cite{shectmanetal1996})
for both quantities following the procedure of JMB98. The error bars
shown in these figures are simply the standard deviations for
identical analyses of the 60 mock samples. We have also used the
bootstrap method to estimate the errors, and found that the bootstrap
errors are typically $30\sim 50 \%$ smaller than the errors determined
from the mock samples. We prefer to adopt the mock errors, because
they have adequately included the sample-to-sample variations
(i.e. cosmic variances). In addition, the mock errors are not
sensitive to the bias model used; the mock samples constructed from
the dark matter particles give similar error estimates (cf. JMB98).
The velocity dispersion $\sigma_{12}(k)$ is nearly a constant from
$0.5$ to $3\mpci$. The average of $\sigma_{12}(k)$ for this $k$ range
is $510\pm 70 \kms$. This error bar for the averaged $\sigma_{12}(k)$
is only slightly smaller than the typical error $\sim 90\kms $ of each
individual $k$ bin, since the errors of individual $k$ bins are
correlated. Our result for $\sigma_{12}(k)$ is in good agreement with
our previous determination $570\pm 80\kms$ at a separation of
$r=1\mpc$ based on a fit to the two-point correlation function
(JBM98). A reasonable change for the $\beta$ value from 0.4 to 0.6
based on the cluster abundance observations (e.g. \cite{ks1996}) leads
to a change of $5\sim 10\%$ in $\sigma_{12}(k)$. The south (dashed
lines) and north (dotted lines) subsamples give a result consistent
with the whole sample within the statistical uncertainty.

Our measurement for the power spectrum $P^R(k)$ is presented in Figure
4. The spectrum is approximately a power law for the range of scales that
the LCRS can explore. It decreases with $k$ approximately as $\propto
k^{-1.7}$ for $0.1\ls k \ls 4 \mpci$.
We consider this result very reliable, because we have tested our
method of a direct Fourier transform extensively with mock samples. In
our tests we found that this method is superior to using other types
of window functions, e.g. spherical ones, especially when the aim is
to obtain unbiased estimates of $\sigma_{12}$ and $P(k)$. These
results are qualitatively in agreement with the real space power
spectrum of the APM survey (\cite{baugh1994}). The shoulder at
$k\approx 0.4 \mpci$ apparent in Figure 4 appeared also in the APM
results (their Figure 11) at exactly the same scale, though the error
bars of the LCRS are much larger. We also note that the power-law
index of $P(k)$ for $0.4\ls k \ls 4 \mpci$ is $\approx -1.2$ compared
with $\approx -1.3$ of the APM survey.

\section{Mock samples and test of the method}
As we discussed in Section 2, there are several possible sources which
could introduce errors to our measured redshift power spectrum and to
our fitted quantities. The use of the finite bins and the cutoffs in
the radial and projected separations may lead to a certain systematic
bias in the measured power spectrum. When we write down equation
(\ref{eq1}), we implicitly make the distant observer assumption. There
are also statistical uncertainties in the power spectrum which are
related to the cosmic variance and the Poisson error inherent to a
redshift survey. All these uncertainties, systematic or statistical,
are however difficult to model with an analytical method. Fortunately,
they can be easily quantified with the help of the mock samples.

We will use the mock samples of JMB98 which were produced from a set
of CDM N-body simulations of 2 million particles. Three flat CDM
models were considered: one is the once standard CDM model of the
density parameter $\Omega_0=1$ and the normalization $\sigma_8=0.62$,
and the other two are low-density flat CDM universes with
parameters $(\Omega_0,\sigma_8)=$ $(0.2,1)$ and $(0.3,1)$
respectively. The known selection effects of the survey,
including the sky boundaries, the limiting magnitudes and the redshift
selection function, the observed rate due to the limited number of
fibers, and the missing galaxies due to the fiber mechanical
collision, were properly simulated in the mock samples.
Because the CDM models were found to have too steep a two-point
correlation and too large a pairwise velocity dispersion compared to
the LCRS result, JMB98 have introduced a simple bias model which under 
weights the cluster regions in the dark matter models. The mock
samples based on this bias model yielded much better matches to the
LCRS observation, and the model of $(\Omega_0,\sigma_8)=(0.2,1)$ was
shown to be consistent with the observations.

To test our statistical method, we will use two types of mock samples
of the cluster weighted $\Omega=0.2$ sample.  To test if there is any
systematic bias with the method, we will use mock samples which
incorporate all selection effects except the fiber collision. The
reason why we exclude the fiber collision effect is because it is not
possible to include this effect in a 3-D simulation sample, as the results
of the 3-D sample must be compared with those of the mock samples in the
test. To quantify the statistical errors of our measured quantities,
we will use the mock samples based on the cluster-weighted
$\Omega_0=0.2$ sample with the fiber collision effect. For these mock
samples, we will include or exclude the fiber collision effect in
order to measure the systematic effect of the fiber collision.

Figure 5 presents our determination of the damping function in the
cluster-weighted mock samples of $\Omega_0=0.2$ according to Equation
(\ref{eq7}). The right panel shows the result for one mock sample
which is randomly chosen. The damping function can be approximated by
a Lorentz function (the solid line) within the scatter as in the LCRS
observation. If we increase the number of mock samples to 60 and thus
reduce the noise, the mean damping function decreases slightly faster
than the Lorentz form at $k\mu\sigma_v\gs 3 $ (left panel). This
behavior was also found by JB2001 for the same model who measured the
damping function by FFT for the full simulations (thus achieving
higher accuracy). In fact, the mean damping function is not only
qualitatively but also quantitatively in good agreement with the
results of JB2001. This test indicates that our determinations for the
damping function are not systematically biased, though there still
exist considerable scatter for the LCRS. With the upcoming large
surveys 2dF and SDSS, we expect that the uncertainties of the redshift
power spectrum will be considerably reduced and it will become
practical to measure a damping function which can be used to
discriminate between theoretical models. For the LCRS, however, the
Lorentz form is still sufficiently accurate for a description of the
damping function.

Using the Lorentz form for the damping function, we determine the
pairwise velocity dispersion $\sigma_{12}(k)$ and the real-space power
spectrum $P^R(k)$ for the mock samples of the $\Omega_0=0.2$ model in
the same way as for the real catalog LCRS. These two quantities are
measured for each of the 60 samples, and the mean values and the
expected $1\sigma$ errors of the mean values are plotted in Figure 6
(symbols with error bars). The agreement is quite satisfactory with
the same two quantities measured from the true particle velocity and
positions in the simulation (solid lines, $k=1/r$ for the pairwise
velocity dispersion). The plots in Figure 6 demonstrate that this
method to estimate $\sigma_{12}(k)$ and $P^R(k)$ works well, and does
not suffer from a strong bias. There does exist some small systematic
bias, as appeared in the estimated velocity dispersion
$\sigma_{12}(k)$ at $k\approx 0.3 \mpci$ and $k\approx 1.5\mpci$,
which is likely caused by the sharp cutoffs in the radial and
projected separations and can be removed by estimating RTPCF to larger
separations.

Next we compare our statistical results with the predictions of the
CDM models using the cluster-weighted bias. The results are shown in
Figure 7. For the model with $\Omega_0=1$, we have shifted the power
spectrum by the factor $1/\sigma_8^2$, assuming a linear bias of
$1/\sigma_8$.  The power spectra are consistent with the LCRS
observation for the range of scales explored by the survey, but the
velocity dispersions of the CDM models are still systematically higher
than the observation. Only the CDM model with the density parameter
$\Omega_0=0.2$ is consistent with the observations. Its $1\sigma$
lower limit actually coincides with the LCRS result. Just as in JMB98
we may conclude that the flat, low-density models give a better fit to
the data than high-density CDM models. The cluster-weighted bias
improves the agreement between models and data remarkably. From Fig.6
we can see that without the cluster-weighted bias (dotted lines) the
velocity dispersion would increase by roughly $100\kms$, and the power
spectrum in real space would be about 2 times higher. Similar changes
could be seen in Fig.7, if the cluster-weighted bias were removed from
the model calculations.

In a previous paper (JB2001) we have shown that the bias model using
primordial density peaks gives even higher values for $\sigma_{12}(k)$
and $P^R(k)$, since the clusters have more weight than the dark matter
particles in that scheme. These arguments carry over to the analysis
of the LCRS data. It seems that a preference for the cluster-weighted
bias emerges from our analysis.

\section{Discussions and conclusions}

Here we want to comment briefly on some previous work on the
determination of the pairwise velocity dispersion $\sigma_{12}$ for
the LCRS. In JMB98 $\sigma_{12}$ has been determined based on the
traditional method of fitting the RTPCF. A value of
$\sigma_{12}=570\pm 80 \kms$ at $r_p=1\mpc$ has been obtained, but
figure 1 of JMB98 shows that this quantity is slightly smaller (about
450 to 500 $\kms$) at smaller ($r_p\ls 0.3\mpc$) or larger ($r_p\gs
3\mpc$) separations. Since some difference between the quantities in
Fourier space and coordinate spaces is expected, we regard our result
in this paper, $\sigma_{12}=510\pm 70\kms$ around $k=1\mpci$, to be in
excellent agreement with the result of JMB98 as well as the recent
results of Ratcliffe et al. (1998). The value is larger than that
Landy et al (1998) obtained based on an power spectrum analysis of the
LCRS who ignored the infall effect and considered $P^s(k,\mu)$ at
$\mu=(0,1)$ only.  In addition, the estimate of the power spectrum
$P^R(k)$ for the LCRS as a direct derivation from the redshift power
spectrum seems a reliable result in the interval between $0.1\mpci$
and $3\mpci$. We suspect that guesses at $P^R(k)$ for larger scales
(smaller $k$) suffer from the sample variation of the LCRS.

We reach the following conclusions:
\begin{itemize}

\item The power spectrum $P^R(k)$, and the pairwise velocity
dispersion $\sigma_{12}(k)$ can be estimated reliably from the
redshift power spectrum of the LCRS. An assumed form for the
distribution function of the pairwise velocity is not needed, although
it turns out that the damping function fits reasonably well to the
Lorentz form, over an interval in $k$ from $0.1\mpci$ to $3\mpci$.
The Lorentz form for the damping function is the Fourier transform of
the exponential distribution of the pair velocities. The fact that
over a range of the scales the Lorentz form is a satisfactory
approximation shows that in a corresponding range of $r$-scales, the
exponential distribution is a good model. A value for $\beta$ must be
assumed, but it has been reasonably determined by other observations.

\item The last values are $\sigma_{12}(k)=510\pm70\kms$ from $k=0.5$ to 
$3\mpci$, in agreement with JMB98; $P^R(k)\propto k^{-1.7}$ for
$k=0.1$ to $4\mpci$.

\item Very important is the extensive use of mock samples which are
constructed from a large set of high resolution simulations. They
allow us to estimate the errors and the statistical significance of
the results reliably, despite the intricate way which various error
producing effects interact.

\item We find reasonable fits to these measured quantities only in
flat, low-density CDM models with a cluster-weighted bias. This
antibias model--the number of galaxies per unit mass in a massive
cluster is proportional to $M^\alpha$ ($M$ is the cluster mass) --
suggested in JMB98 obviously deserves attention in future work. It
should be pointed out that the parameter $\alpha=-0.08$ used in this
paper corresponds to a particular selection of galaxies: the LCRS
galaxies. For other mix of galaxies (i.e., other observational
criteria), $\alpha$ could be different. In our recent work (in
preparation), we found the bias model of $\alpha=-0.25$ can
successfully account for the clustering of IRAS galaxies.

\item The method applied here would be more powerful in samples larger
than the LCRS. The power spectrum $P^R(k)$, the $\beta$-value,
pairwise velocity dispersion, and the damping function can all be
determined simultaneously, if the data are good enough.

\end{itemize}

\acknowledgments

We are grateful to Yasushi Suto for the hospitality extended to us at
the physics department of Tokyo university where most of the
computation was completed. J.Y.P. gratefully acknowledges the receipt
of a NAO COE foreign research fellowship. The work is supported in
part by the One-Hundred-Talent Program, by NKBRSF(G19990754) and by
NSFC to Y.P.J., and by SFB375 to G.B..

\begin{figure}
\epsscale{1.0} \plotone{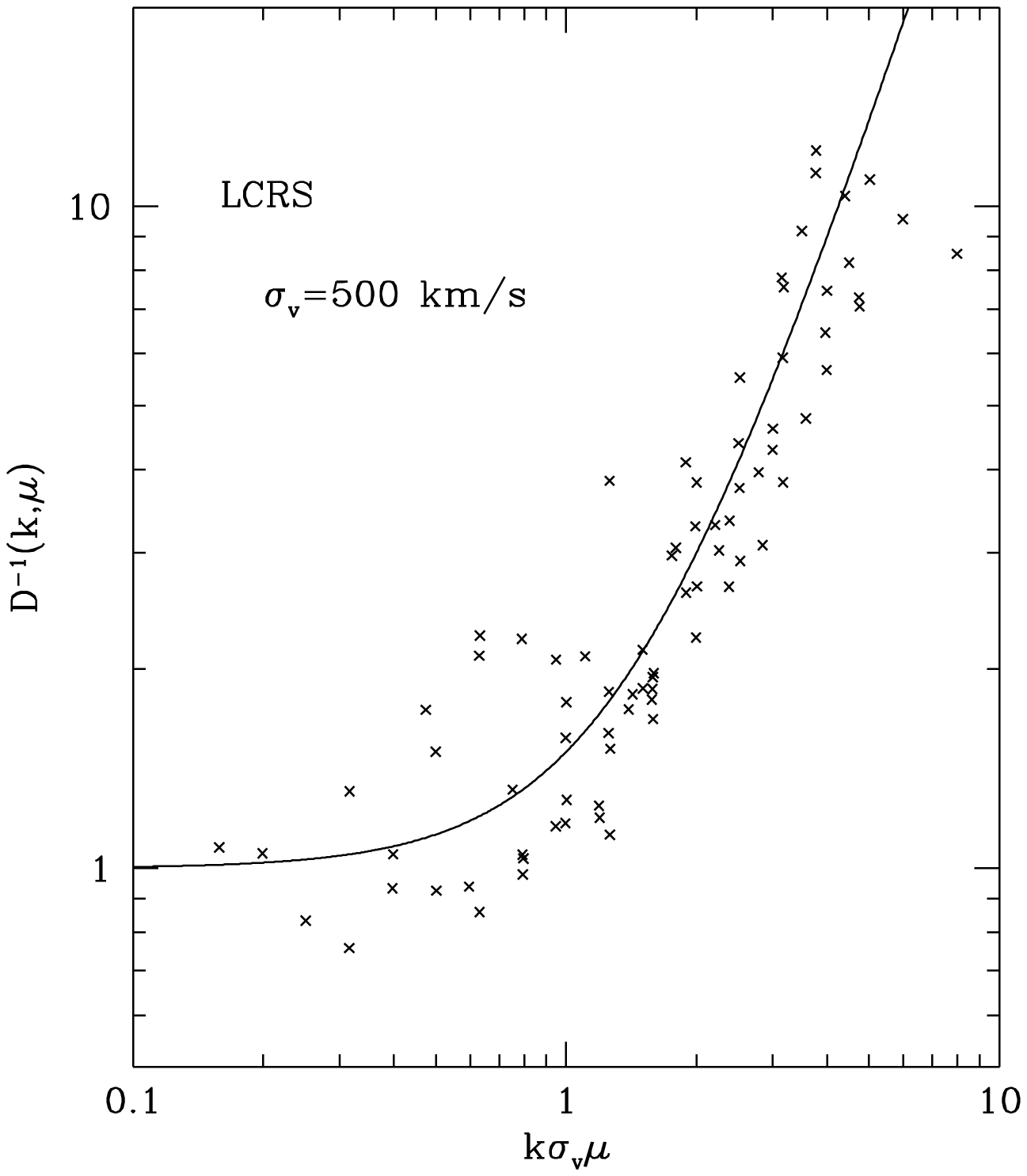}
\caption{The inverse of the damping function $D(k,\mu)$ of the Las Campanas Redshift
  Survey, plotted as a function of $k\mu\sigma_v$. In this plot, we
  have taken $\beta=0.5$ and $\sigma_v=500 \kms$.
}\label{fig1}\end{figure}

\begin{figure}
\epsscale{1.0} \plotone{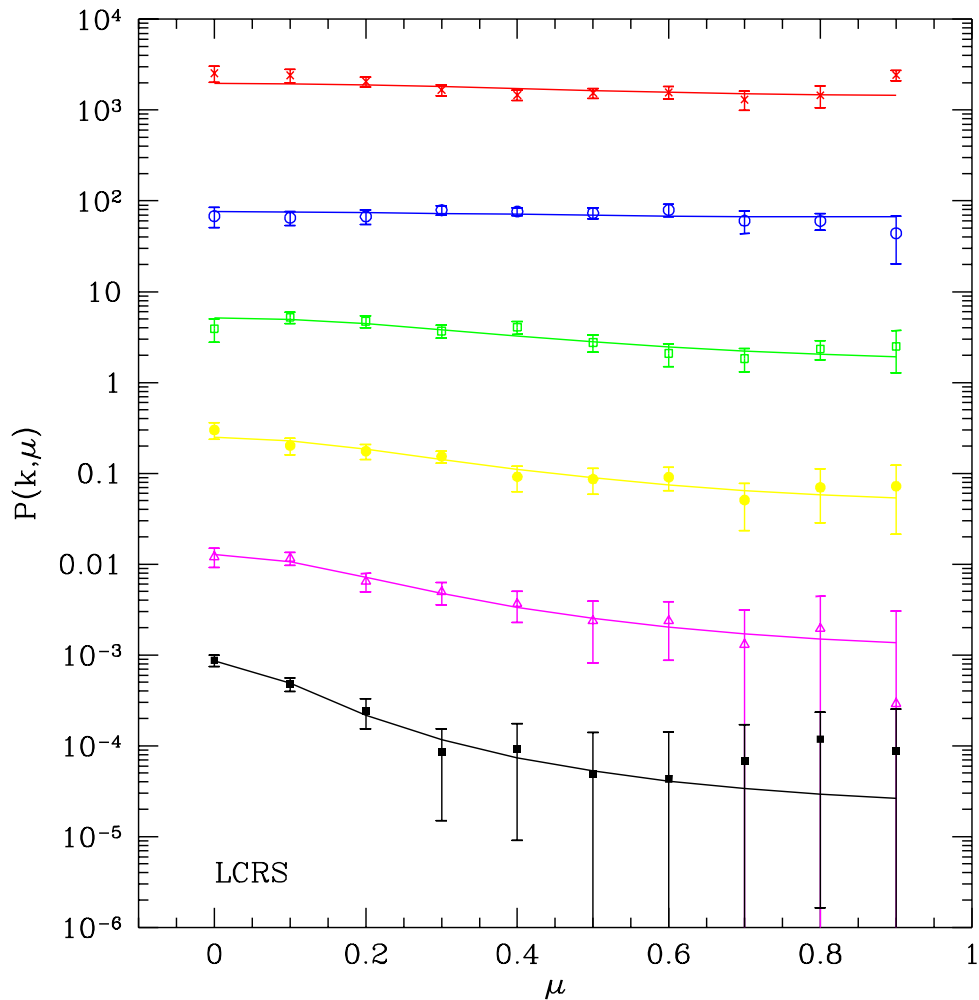}
\caption{The redshift power spectrum $P^S(k,\mu)$ of the Las Campanas
  Redshift Survey. The solid lines are the best-fit curves for free
  parameters $\sigma_v$ and $P^R(k)$, assuming the Lorentz form for
  the damping function. Different colors correspond to
different values of $k$. From top to bottom, $k= 0.25$ to $2.5\mpci$
at $\Delta \lg k=0.2$.
}\label{fig2}\end{figure}

\begin{figure}
\epsscale{1.0} \plotone{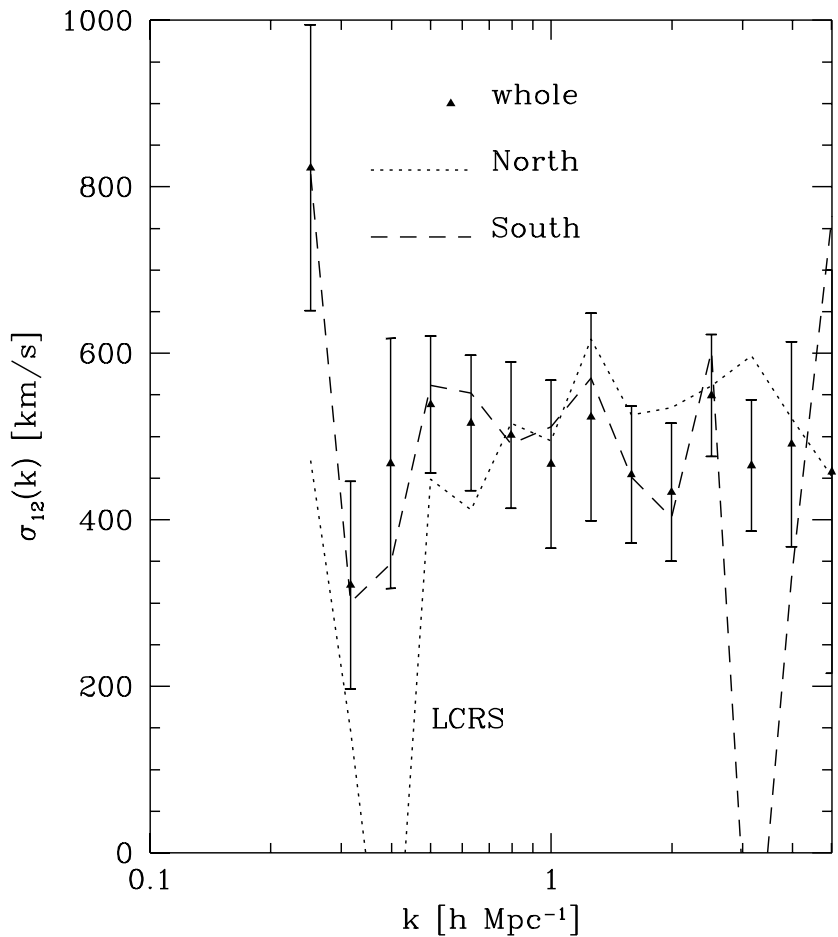}
\caption{The pairwise velocity dispersion determined from the redshift
power spectrum for the Las Campanas Redshift Survey. The whole
sample, the south subsample, and the north subsample are represented
by the triangles (with error bars), the dashed and the dotted lines
(without error bars) respectively.}\label{fig3}\end{figure}

\begin{figure}
\epsscale{1.0} \plotone{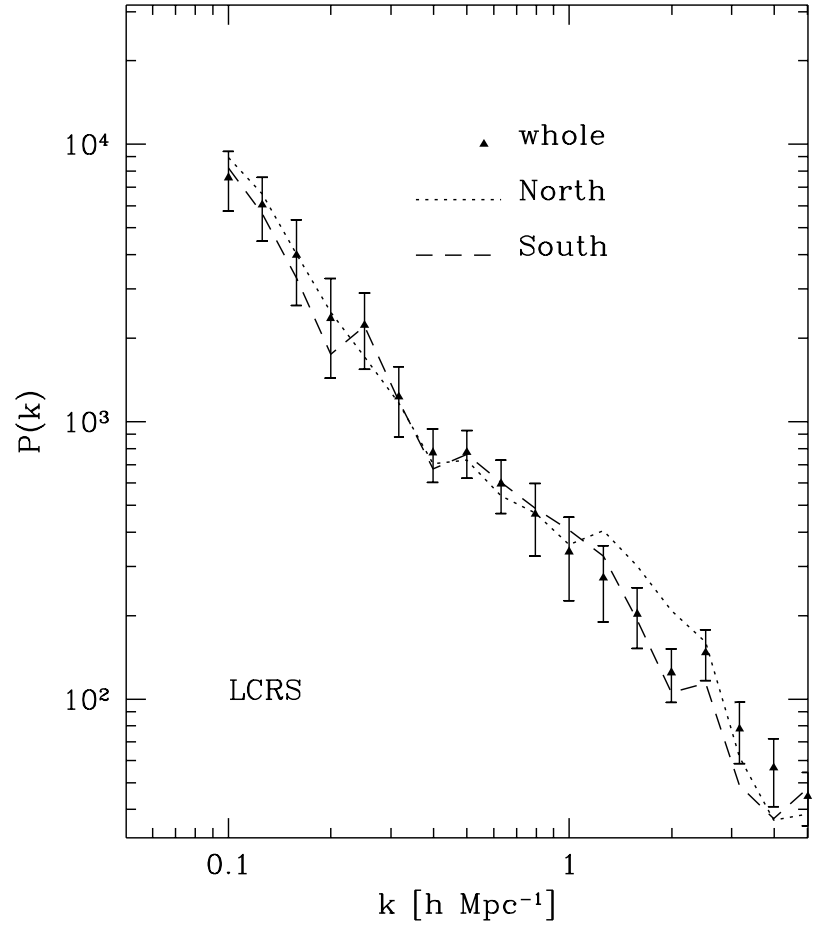}
\caption{The real space power spectrum determined from the redshift
power spectrum for the Las Campanas Redshift Survey. The whole
sample, the south subsample, and the north subsample are represented
by the triangles (with error bars), the dashed and the dotted lines
(without error bars) respectively. 
}\label{fig4}\end{figure}

\begin{figure}
\epsscale{1.0} \plotone{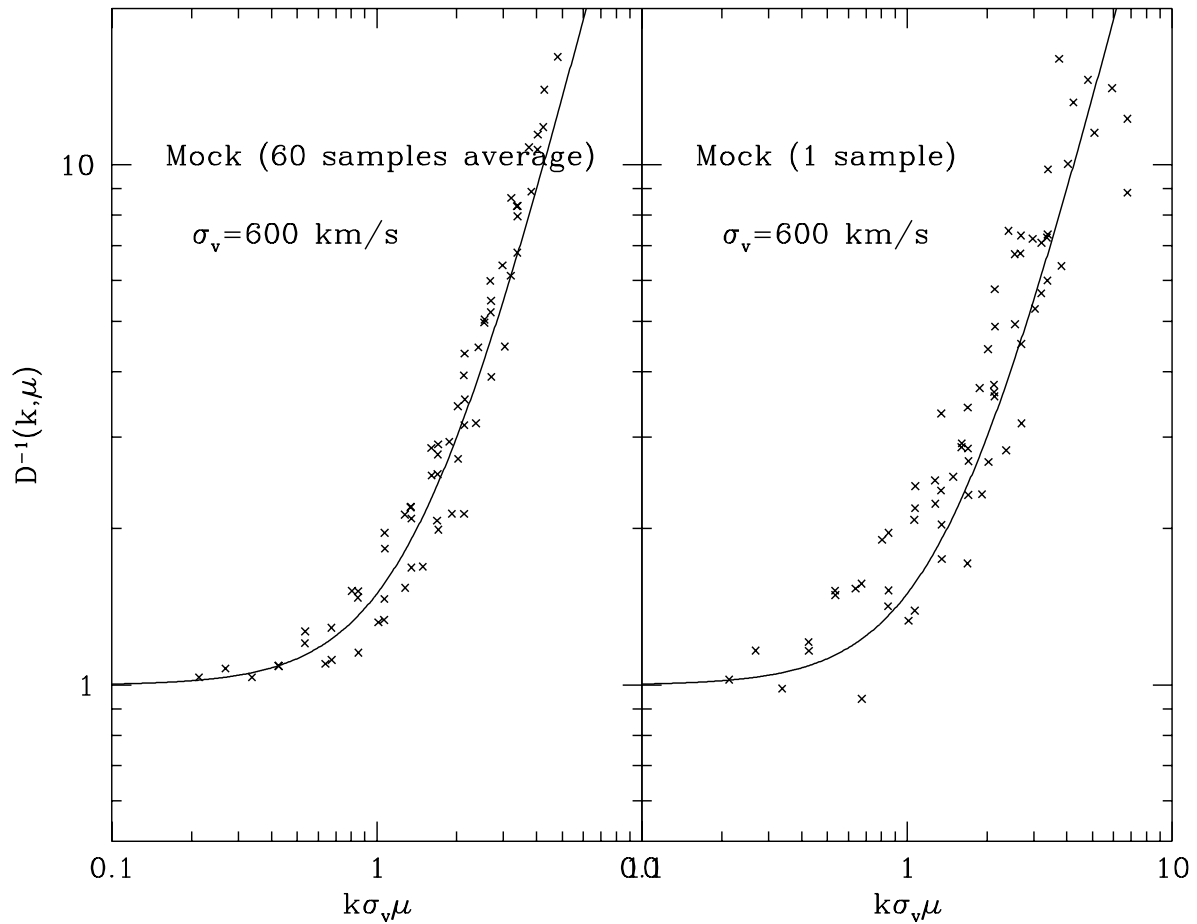}
\caption{The inverse of the damping function $D(k,\mu)$ of the cluster-weight biased mock
  samples with $\Omega_0=0.2$, plotted as a function of
  $k\mu\sigma_v$. In this plot, we have set $\sigma_v=600 \kms$.
  }\label{fig5}\end{figure}

\begin{figure}
\epsscale{1.0} \plotone{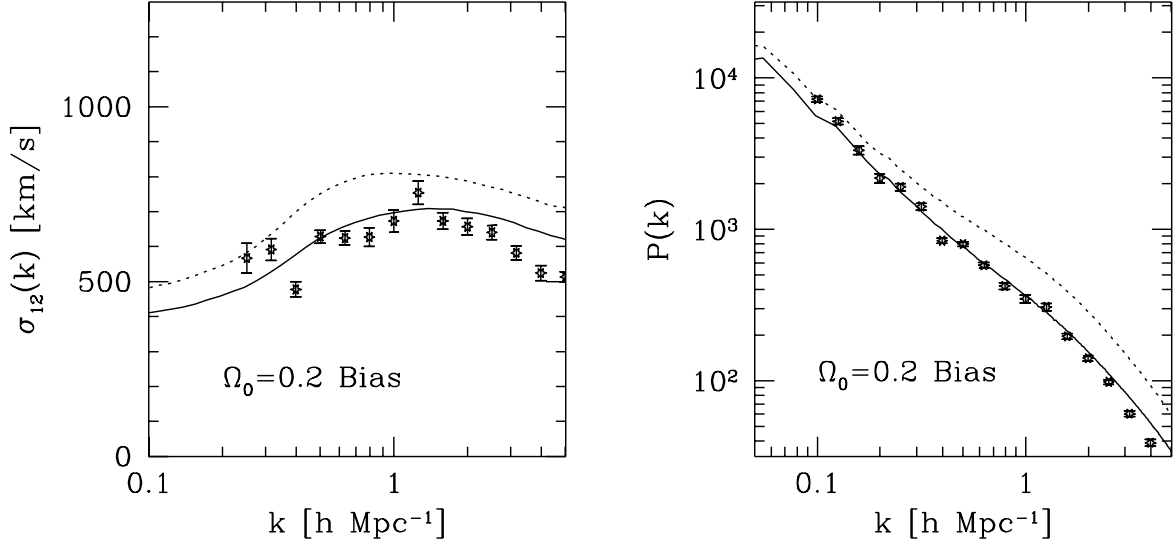}

\caption{The pairwise velocity dispersion and real-space power
  spectrum determined for the 60 mock samples. The error bar is the
  expected $1\sigma$ error of the mean value of the 60 samples. The solid
  lines are the pairwise velocity dispersion measured from the
  peculiar velocity (assuming $k=1/r$ for the plot) and the real-space
  power spectrum from the particle spatial coordinates. Similar to the 
  solid lines, the dashed ones are measured from the full simulations
  but without the cluster-weighted bias.
}\label{fig6}\end{figure}

\begin{figure}
\epsscale{1.0} \plotone{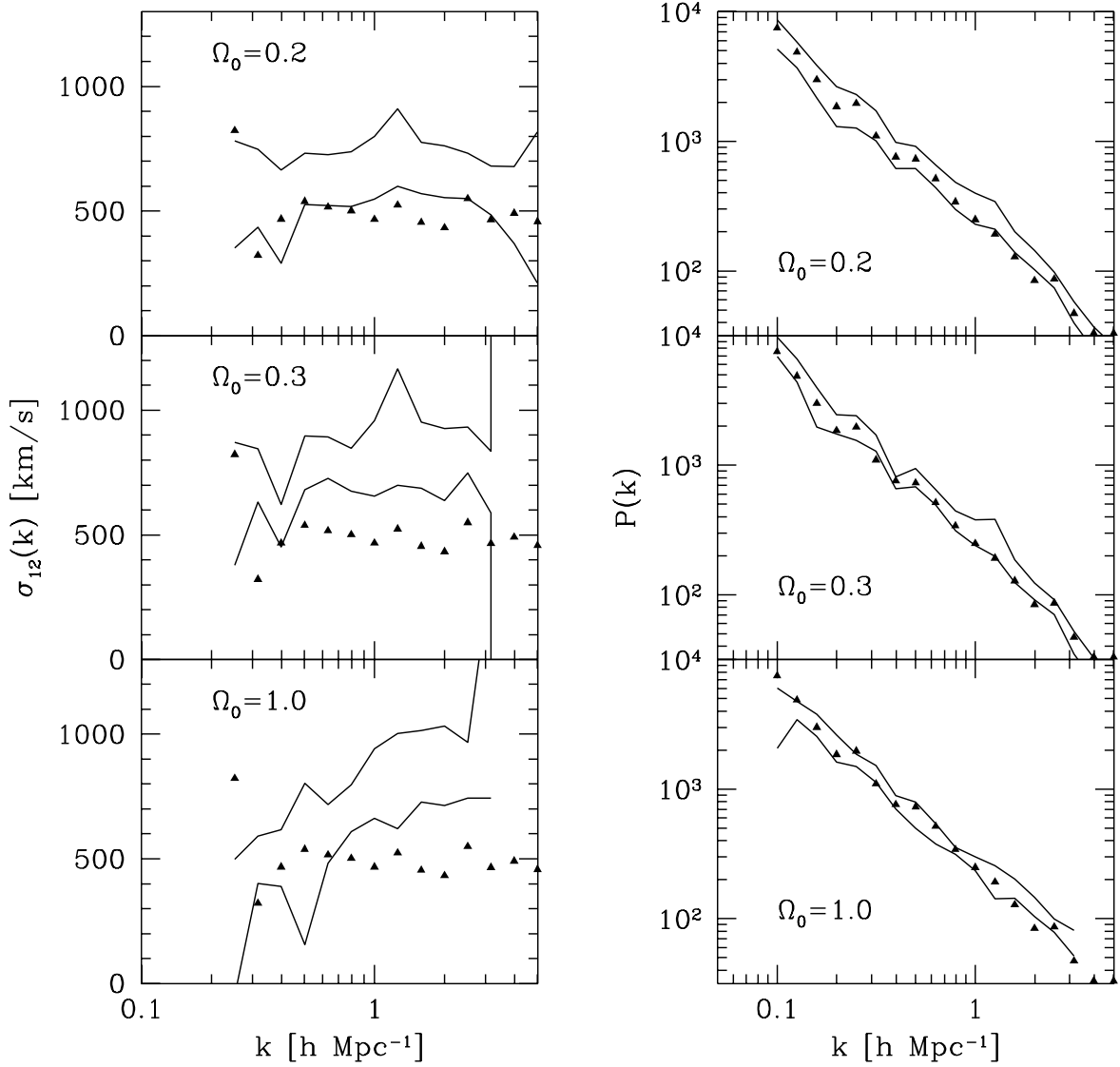}
\caption{The pairwise velocity dispersion and real-space power
  spectrum determined for the Las Campanas Redshift Survey, compared
  to three CDM models with the cluster-weighted bias. The
  triangles are the results of the LCRS, and the lines are the
  $1\sigma$ upper and lower limits of the CDM models. 
}\label{fig7}\end{figure}

\end{document}